\documentclass[aps,prl,twocolumn,superscriptaddress]{revtex4-1}

\usepackage[pdftex]{graphicx}

\begin{document}

\title{Reply to ``Comment on \emph{``Precision global measurements of London penetration depth in Fe(Te$_{0.58}$Se$_{0.42}$)''}''}

\author{K.~Cho}
\affiliation{The Ames Laboratory, Ames, IA 50011, USA}
\affiliation{Department of Physics \& Astronomy, Iowa State University, Ames, IA 50011, USA}

\author{H.~Kim}
\affiliation{The Ames Laboratory, Ames, IA 50011, USA}
\affiliation{Department of Physics \& Astronomy, Iowa State University, Ames, IA 50011, USA}

\author{M.~A.~Tanatar}
\affiliation{The Ames Laboratory, Ames, IA 50011, USA}
\affiliation{Department of Physics \& Astronomy, Iowa State University, Ames, IA 50011, USA}

\author{R.~Prozorov}
\email[Corresponding author: ]{prozorov@ameslab.gov}
\affiliation{The Ames Laboratory, Ames, IA 50011, USA}
\affiliation{Department of Physics \& Astronomy, Iowa State University, Ames, IA 50011, USA}

\date{15 July 2012}

\begin{abstract}

We reply to the Comment by T.~Klein, P.~Rodi\`{e}re and C.~Marcenat \cite{KleinComment}, on our paper, \emph{``Precision global measurements of London penetration depth in FeTe$_{0.58}$Se$_{0.42}$''}, Phys. Rev. B \textbf{84}, 174502 (2011). Our work was motivated by Klein \emph{et. al}, \emph{``Thermodynamic phase diagram of Fe(Se$_{0.5}$Te$_{0.5}$) single crystals in fields up to 28 Tesla''}, Phys. Rev. B \textbf{82}, 184506 (2010). In their paper, Klein \emph{et al.} have attributed a factor of five difference in the value of the London penetration depth obtained from their tunnel diode resonator (TDR) measurements and calculated from the ``field of first penetration'' to the surface roughness, although they have not verified it experimentally. In our paper, we have studied the effects of deliberately introduced surface roughness and found that its effects are minor and cannot be responsible for the difference of such magnitude. Instead, we suggest that the value of the ``field of first penetration'' measured with Hall - arrays cannot be used to extract a true lower critical field due to several reasons outlined in our Reply. We emphasize that the accuracy of the calibration procedure of the TDR technique has been carefully verified in several prior studies and our work on FeTe$_{0.58}$Se$_{0.42}$ further confirms it. We show that in their Comment, Klein \emph{et. al} use wrong arguments of the universal behavior of the superfluid density in the gapless limit, because it is inapplicable for the multi - band superconductors. We also discuss the applicability of the clean - limit $\gamma-$ model and the influence of the disorder on the obtained results.
\end{abstract}

\pacs{74.70.Xa,74.20.Rp,74.62.En}


\maketitle
In their paper on the properties of Fe(Te$_{0.5}$,Se$_{0.5}$) superconductor \cite{Klein2010PRB}, Klein \emph{et al.} have estimated the ``lower critical field'' from the measurements of the ``field of first penetration'' measured by using miniature Hall - probe array and they also measured a change in the London penetration depth, $\Delta \lambda (T)$, using tunnel diode resonator (TDR) technique. The two measurements have produced conflicting results, which Klein \emph{et al.} have attributed to a \emph{factor of five difference} between TDR calibration and the ``real'' $\Delta \lambda$ \emph{due to the surface roughness}. However, Klein \emph{et al.} have newer verified their conjecture experimentally.

The application of the tunnel - diode resonator (TDR) technique to study superconductors has been in development for the past 20 years (see e.g., Refs.~\cite{Prozorov2006SST,Prozorov2011RPP}) and its calibration procedure has been firmly established and verified on various systems starting from the original works on Nb \cite{Prozorov2000PRB} and the cuprates \cite{Prozorov2000a}. More recently we carried out a joint study by TDR, microwave cavity perturbation technique and muon - spin rotation ($\mu$SR) measurements obtaining close results on Ba(Fe,Co)$_2$As$_2$ (BaCo122) pnictide superconductor \cite{uSR_MW_TDR_FeCo122_2012}. Finally, we recently measured the same sample by using \emph{local} scanning SQUID and TDR and obtained excellent agreement between the two techniques \cite{Lippman2012}. We found that it was very important to study the same physical samples to arrive at this conclusion.

In their Comment, Klein \emph{et al.} support their conjecture of the influence of surface roughness by citing a study of a nodal superconductor, KFe$_2$As$_2$, by Hashimoto \emph{et al.} \cite{Hashimoto2010PRB_KFe2As2} in which the Authors have found the difference in the penetration depth only \emph{by a factor of two}. We note, however, that these were two physically different samples with different temperature - dependent $\Delta \lambda (T)$. Yet, even this difference has resulted only in a factor of two, which is very far from a factor of five claimed by Klein \emph{et al.}). We also note the earlier study of the superfluid density in (Ba$_{1-x}$K$_x$)Fe$_2$As$_2$ by the same Kyoto group, in which the Authors explained the difference between different samples to be due to different scattering rates \cite{Hashimoto2009BaK122}.

Puzzled by Klein \emph{et al.} claim of such a profound effect of the surface roughness, we studied its effect experimentally. We used \emph{the same crystal} to deliberately damage and re-measure $\lambda(T)$ after each damage. The sample was cut with a razor blade, -  a technique commonly employed in handling soft samples of Fe - based superconductors. Starting with a clean cut, we then deliberately damaged the edges (the sample has lost about 10~$\%$ of its volume after this step) and, finally, removed the damage by a new clean cut. London penetration depth was measured after each step. It was found that the penetration depths measured before and after razor-blade damaging didn't show any significant change and we concluded that roughness plays only a minor role, even for a very strong damage \cite{Cho2011PRB_FeTeSe} invalidating Klein \emph{et al.} claim \cite{KleinComment}.

In their comment \cite{KleinComment} Klein \emph{et al.} state that \emph{``Cho et al. completely ignored the large dispersion in the previously published TDO data.''}. This statement is incorrect and, as we show below, different samples of Fe - based superconductors could well be in different scattering regimes. Indeed, there can be quite significant difference in the measured properties of nominally similar samples produced by different groups and at different times. Our own measurements of BaCo122 material is a good evidence of that, see Ref.~\onlinecite{Prozorov2011RPP} for details. Similarly, there is a difference between previous studies of FeTe$_{0.5}$Se$_{0.5}$ superconductor \cite{Kim2010c, Serafin2010PRB, Klein2010PRB} and a more recent study \cite{Cho2011PRB_FeTeSe}. As shown in Ref.~\onlinecite{Cho2011PRB_FeTeSe} it has nothing to do with the surface roughness, but reflects a real compositional difference between studied samples.

In our opinion, the discrepancy found in Ref.~\cite{Klein2010PRB} could be attributed to a misinterpretation of what Klein \emph{et al.} call the ``lower critical field'', $H_{c1}$. First, Klein \emph{et al.} use 8 x 8 $\mu$m Hall probe arrays spaced at the same distance of 8 $\mu$m. Detecting real $H_{c1}$ would require a spatial resolution better than the London penetration depth (about 0.5  $\mu$m in Fe(Te,Se)\cite{Kim2010c}), which is clearly impossible with their technique that at best resolves about 10 $\mu$m. Second, they use analysis based on Brandt model developed for \emph{pinning – free} superconductor where the field of the first flux penetration is determined by the geometric barrier without considering surface roughness (on the scale of the coherence length!) and ignoring possible Bean - Livingston barrier \cite{Brandt1999}. There is absolutely no evidence for a geometric barrier - limited behavior in Fe(Te,Se) crystals. To the contrary, there is a considerable vortex pinning \cite{FeTeSePinning}. Ironically, at the same, time Klein \emph{et al.} use surface roughness argument to stretch the TDR data to fit this ill – defined ``$H_{c1}$''.

The major part of Klein \emph{et al.} comment \cite{KleinComment} questions our use of a clean - limit model for the description of the superfluid density in FeTe$_{0.58}$Se$_{0.42}$. Indeed, as was found already in the early studies, the low - temperature exponents of the order of two might be due to pair - breaking scattering \cite{Hashimoto2009BaK122,Gordon2009,Gordon2010}. However, it does not imply that the sample used in Ref.~\onlinecite{Cho2011PRB_FeTeSe} is in the dirty limit. At a moderate pair - breaking, the lower temperature range is affected first (similar to $d-$wave cuprates where $T-$linear behavior changes to $T^2$ at the low temperatures first \cite{Hirschfeld1993PRB}). The main point of Klein \emph{et al.} comment is their Fig.~1, where they use ``universal'' behavior of the superfluid density \emph{in the gapless limit}, $\rho(t) \sim (1-t^2)$, where $t=T/T_c$ in the \emph{entire temperature interval} apparently following our earlier works (see, e.g., Eq.~45 in \onlinecite{Prozorov2011RPP}). Unfortunately for Klein \emph{et al.}, this universal behavior only holds for a single - band superconductor, albeit with arbitrary anisotropy of the Fermi surface and of the superconducting gap, where one could use the conjecture of the separation of temperature and angular dependence of the order parameter ($\Omega^2-$ approximation) with proper normalization, see Eqs.~(15,16) in Ref.~\onlinecite{Prozorov2011RPP}. For the two - band case, we had to turn to a full Eilenberger scheme (the $\gamma-$model \cite{Kogan2009gamma}), see Sec.~3.2 in Ref.~\onlinecite{Prozorov2011RPP}, which no longer uses $\Omega^2-$ approximation and, instead, uses full interaction matrix of three coupling constants. Similarly, we could not use $\Omega^2-$ approximation to analyze the $H_{c2}$ problem in a multi - band case \cite{Kogan2011Hc2}. Iron - based superconductors are multi - band, multi - gap materials and universal $\rho(t) \sim (1-t^2)$ behavior is not expected to work in the full temperature range, so the apparent agreement with this function with the data in Klein \emph{et al.} Comment is a mere coincidence.

To study the effect of pair - breaking scattering in a multi - band scenario one has to solve a complete microscopic problem with the interaction matrix and it does not lead to a $\rho(t) \sim (1-t^2)$ behavior. There will be six new scattering parameters (magnetic and non - magnetic for two intra - band and one inter - band channels), which are unknown. In Fe - based superconductors pair - breaking scattering rates can vary dramatically. For example, the effect on BaCo122 system is discussed in Ref.~\onlinecite{Prozorov2011RPP}. Moreover, scattering rates may be anisotropic \cite{HusseyAnisotropicScattering2007} and temperature - dependent \cite{Maeda2011}. In fact, the latter work studied the same material, Fe(Te,Se), as the subject of this Reply and Comment. The Authors used microwave surface impedance measurements to conclude that the temperature - dependence of the scattering rate can be so significant that it may result in a crossover from dirty to clean regimes within the superconducting domain \cite{Maeda2011}. Indeed, no universality of $\rho(t)$ is expected in such case.

Finally, returning to the claim by Klein \emph{et al.} that we conduct our analysis \emph{``... completely ignoring the large dispersion in the previously published TDO data.''}. Our answer is the following. We believe that the earlier data on Fe(Te,Se) crystals (including those of Klein \emph{et al.}) \cite{Kim2010c,Klein2010PRB,Maeda2011} represent the dirtier cases, whereas the sample in our more recent study \cite{Cho2011PRB_FeTeSe} is closer to the clean behavior. Therefore, fitting the superfluid density with a clean - limit self - consistent two - gap model in the full temperature range is still justified at least in Ref.~\onlinecite{Cho2011PRB_FeTeSe} and results in a meaningful estimate of the interaction matrix and of the derived temperature - dependent superconducting gaps. On the other hand, the \emph{five - fold difference} between ``$H_{c1}$'' estimated by Klein \emph{et al.} and their TDR data collected on the same crystal supports our conclusion that the quantity they measure is not a true $H_{c1}$.

We thank V. G. Kogan for discussions. The work at Ames was supported by the U.S. Department of Energy, Office of Basic Energy Sciences, Division of Materials Sciences and Engineering under contract No. DE-AC02-07CH11358.

\end{document}